\documentclass[twocolumn,aps,showpacs,prr,amsmath,amssymb,floatfix]{revtex4-2}

\usepackage{amsmath,amssymb,color}

\usepackage{bm}
\usepackage{epsfig}
\usepackage{psfrag}
\usepackage[latin1]{inputenc}
\usepackage[english]{babel}
\usepackage{amsfonts}
\usepackage{amsmath}
\usepackage{upgreek}
\usepackage[unicode,hidelinks]{hyperref}
\usepackage[dvipsnames]{xcolor}

\DeclareSymbolFont{bbold}{U}{bbold}{m}{n}
\DeclareSymbolFontAlphabet{\mathbbold}{bbold}

\newcommand{\e}{{\rm e}}
\newcommand{\ket}[1]{| #1 \rangle }
\newcommand{\bra}[1]{\langle #1  |}

\newcommand{\ex}[1]{\langle #1 \rangle}	
\newcommand{\up}{\uparrow}	
\newcommand{\down}{\downarrow}	
\newcommand{\matt}[4]{ \left(\begin{matrix}#1 &#2\\#3&#4\end{matrix}\right)}

\begin{document}

\title{
	Transmission Amplitude through a Coulomb blockaded Majorana Wire
	}

\author{
	Matthias Thamm and Bernd Rosenow
	}
\affiliation{
	Institut f\"{u}r Theoretische Physik, Universit\"{a}t
  Leipzig,  Br\"{u}derstrasse 16, 04103 Leipzig, Germany
	}
	
\date{\today}

\begin{abstract}
		We study coherent electronic transport through a Coulomb blockaded
		superconducting Rashba wire in the co-tunneling regime between 
		conductance resonances.  By varying an external Zeeman field the 
		wire can be tuned into a topological regime, where non-local 
		transport through Majorana zero modes is the dominant mechanism.  
		We model coherent transport in the co-tunneling regime by using a 
		scattering matrix formalism, and find that the transmission amplitude 
		has a maximum as a function of Zeeman field, whose height is 
		proportional to the wire length. We relate the transmission amplitude 
		to the Majorana correlation length, and argue that the Zeeman field 
		and length dependence of the transmission amplitude are unique 
		signatures for the presence of Majorana zero modes. 
\end{abstract}
 
\maketitle

\section{Introduction}
		In recent years, Majorana zero modes (MZMs) have attracted much 
		attention as possible candidates for the realization of topologically 
		protected quantum bits \cite{Alicea.2011,Hyart.2013,Clarke.2011}. MZMs  
		can arise as localized zero energy excitations in topological 
		superconductors under suitable conditions \citep{Alicea.2012, 
		Beenakker.2013, Antipov.2018, Schuray.2020}, and many of their predicted experimental 
		signatures have been observed, for instance a zero-bias conductance 
		peak \citep{Mourik.2012,Deng.2012,Das.2012,Finck.2013,Churchill.2013} 
		and  the suppression of the even-odd splitting of Coulomb blockade 
		resonances in the topological phase \citep{Albrecht.2016}.\\
		\indent The topological nature of MZMs manifests itself in  their non-local 
		character \citep{Nayak.2008,Stern.2008}. We study how 
		the non-locality of the electronic state encoded by MZMs can be probed 
		by  phase coherent transport through a  Coulomb blockaded wire with MZMs 
		at its ends \citep{Fu.2010,Landau.2016,Plugge.2016,Vijay.2016,
		Plugge.2017,Hell.2018,Drukier.2018}. By embedding the Majorana wire into 
		the arm of an electron interferometer, the amplitude of coherent 
		transmission through the MZMs can be studied \citep{Whiticar.2020}. In a 
		conductance valley in between Coulomb blockade peaks, the amplitude of the 
		transmission through a Majorana wire is  determined by the magnitude of 
		the wave functions at the ends of the wire. For MZMs, the wave function 
		has a large magnitude $\propto 1/\sqrt{\xi}$ near the wire end, where	$\xi$
		denotes the Majorana correlation length. In contrast, if the transmission 
		is dominated by transport through  extended states, the magnitude of wave 
		functions $\propto 1/\sqrt{L}$ depends on the wire length. Taking into 
		account the decrease of the charging energy with wire length, in a 
		conductance valley one expects an increase of coherent transmission 
		$\propto L$ when entering the topological regime, a robust signature of 
		MZMs which also allows to distinguish them from pseudo-MZMs  
		\citep{Prada.2012,Kells.2012,Cayao.2015,SanJose.2016,Hell.2018, Reeg.2018, Awoga.2019, 
		Vuik.2019}, which can arise in the presence of a soft confinement at the wire ends. 
		Details of the magnetic field dependence follow from the relation 
		$1/\xi = \Delta_{p,\rm ind} /(\hbar v_F)$, where		
		$\Delta_{p,\rm ind}$ is the topological p-wave superconducting gap, 
		 in agreement with results of a recent experiment 
		\citep{Whiticar.2020}.

		\begin{figure}[t!]
			\centering
			\includegraphics[width=8.6cm]{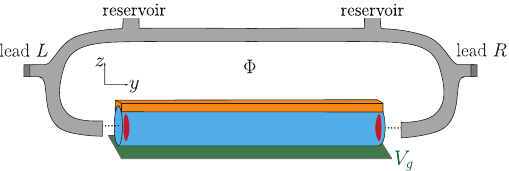}
			\caption{\label{Fig:SetupAAA}Schematic sketch of the Majorana interferometer 
						setup. The lower arm contains the quantum dot consisting of a 
						one dimensional Rashba wire (blue), superconductor (orange), 
						and gate (green). The upper arm of the interferometer is the 
						reference arm which only contains a wire. By varying a flux 
						$\Phi$, the transmission amplitude of electrons tunneling 
						through the dot as a function of the gate voltage can be
						observed \citep{Drukier.2018}. When the wire is tuned to the 
						topological regime by an external Zeeman field, Majorana zero 
						modes (red) are present at the ends. A reservoir is needed to 
						avoid the phase rigidity effect \citep{Aharony.2002}. \vspace{-0.2cm}}
		\end{figure}

\section{Model}
\subsection{Setup}
		In this article, we consider a one-dimensional Rashba wire in proximity to 
		an s-wave superconductor and subject to a perpendicular magnetic field. 
		This system can be tuned into a topological regime where it realizes MZMs 
		at the ends of the wire \citep{Read.2000,Fu.2008,Sau.2010,Lutchyn.2010,
		Oreg.2010,Alicea.2011}. We consider a setup depicted in Fig.~\ref{Fig:SetupAAA} 
		where a wire in the Coulomb blockade regime is tunnel coupled to  leads 
		at each end and embedded into one arm of an Aharonov-Bohm interferometer. 
		Thus, the combination of wire and superconductor acts as a  quantum dot. 
		By adjusting the flux $\Phi$ through the Aharonov-Bohm ring and measuring 
		the conductance oscillations, it is possible to extract the complex 
		transmission amplitude. Here, we focus on the magnitude of the transmission 
		amplitude, which includes phase information in the thermal average over 
		occupations of the dot. For this reason, the transmission amplitude is 
		able to distinguish MZMs from pseudo-MZMs, which is not possible when considering
		conductance measurements only.

\subsection{Transmission amplitude}
		In lowest order interference, the current through the interferometer is given by 
		\begin{equation}
			I(\Phi)  \propto 	2|T_{\rm ref}|^2 
			+ \sum_{\sigma\sigma^\prime} |T_{\sigma\sigma^\prime}|^2 
			+ 2\sum_\sigma {\rm Re}\!\left[\e^{i\Phi}T_{\rm ref}T_{\sigma\sigma}^*\right] 
			\label{Eq:Curr}
		\end{equation}
		where $T_{\rm ref}$ is the transmission amplitude through the reference 
		arm (assumed to be diagonal in spin). Here, $T_{\sigma\sigma'}$ is the	
		transmission amplitude of coherently tunneling electrons with spin quantum number
		$\sigma$, $\sigma'$, which is an entry of the scattering matrix determined by the 
		Mahaux-Weidenm\"{u}ller formula \citep{Mahaux.1968}
		\begin{equation}
			S = 1-2\pi i \left\langle W\,\frac{1}{\varepsilon-H_{\rm eff}
					+i \pi W^\dagger W}\, W^\dagger \right\rangle\ . 
			\label{Eq:Weid}
		\end{equation}
		The brackets denote the thermal average over occupations $\{n_i\}$ of  BdG 
		eigenstates in the dot, defined as $ \ex{\mathcal{O}} 
		= 1/Z\,\sum_{\{n_i\}} \e^{-\beta E(\{n_i\})} \mathcal{O}(\{n_i\})$. The 
		average is performed for fixed total particle number $N_0$, which determines 
		the number parity of occupied BdG levels  $\{n_i\}$. Here, $\varepsilon$ is 
		the energy of incoming electrons, and the effective dot	Hamiltonian 
		$H_{\rm eff}$ and the matrix of dot-lead couplings $W$ with lead index $\alpha$
		and spin $\sigma$ are given by
		\begin{widetext}
		\begin{align}
			H_{\rm eff}&= \matt{{\rm diag}\left[{\varepsilon}_{j}^h(N_0,\{n_i\})
			 \right]_{j=1..j_{\rm max}}}{0}{0}{{\rm diag}
			 \left[{\varepsilon}_{j}^e(N_0,\{n_i\})\right]_{j=1..j_{\rm max}}}\;, 
			 \label{Eq:Heff}\\
			(W)_{\alpha \sigma} 
			 &= \sqrt{\varrho_F} \left({\lambda}^h_{\alpha 1 \sigma}(N_0,\{n_i\}),
			 \ldots,{\lambda}^h_{\alpha j_{\rm max} \sigma}(N_0,\{n_i\}), 
			 {\lambda}^e_{\alpha 1 \sigma}(N_0,\{n_i\})
			 ,\ldots,{\lambda}^e_{\alpha j_{\rm max} \sigma}(N_0,\{n_i\}) \right)\ . 
			 \label{Eq:W}
		\end{align}
		\end{widetext}
		The energies for electron(hole)-like tunneling processes 
		${\varepsilon}_{j}^{e(h)}(N_0,\{n_i\})$ contain both charging energy and  
		single particle energy levels of the wire Hamiltonian. To describe 
		co-tunneling processes, we consider the dot in an initial state 
		$\ket{N_0,\{n_i\}}$. The transmission then occurs via an intermediate 
		state $\ket{N_0\pm1, \{n_i'\}}$, where the allowed occupation numbers 
		$\{n_i'\}$ of the intermediate state deviate from those of the initial 
		state by adding or removing a single Bogolubon, and by adding (removing) one electron 
		charge to the dot. The electron(hole)-like couplings 
		${\lambda}^{e(h)}_{\alpha j \sigma}(N_0,\{n_i\})$ of lead $\alpha$ to  
		level $j$ in the dot are obtained from the overlap 
		$\bra{N_0,\{n_i\};\{\alpha,\sigma\}}H_{\rm tun}\ket{N_0 \pm1, \{n_i'\}}$, 
		with $H_{\rm tun}$ defined in Eq.~\eqref{Eq:tunneling}. 
	
		In the topological regime,  exponentially localized MZMs  occur, for 
		instance at the left end of the wire with the wave function 
		$\chi_{\sigma,L}(y)$ with envelop $\xi^{-1/2} \e^{-y/\xi}$.	 
		In the presence of a small overlap between the left and right MZM, the BdG 
		eigenfunctions are given by $(\chi_{\sigma,L} \pm \chi_{\sigma,R})/\sqrt{2}$. 
		Evaluating Eq.~\eqref{Eq:Weid} to leading order in the dot-lead couplings, one
		finds that the transmission amplitude	through the MZMs is $T_{\sigma\sigma} 
		\sim \chi_{\sigma,L}(y_L)\chi_{\sigma,R}^*(y_R)/(E_c/2)$. 
		Thus, the transmission amplitude provides direct information about the  
		Majorana localization length $\xi$. 
		
\subsection{Hamiltonian}
		We describe the proximitized semiconductor wire by the Hamiltonian
		\begin{equation}
		 \begin{split}
			\mathcal{H}_{\rm wire}&=\tau_z\otimes \left[-\frac{\hbar^2\partial_y^2}{2m^*}
			 \,\sigma_0-\mu\sigma_0-i\hbar\alpha_R\sigma_x\partial_y\right]\\
			 &\phantom{=\;}-E_z\tau_0\otimes\sigma_z+\Delta\tau_x\otimes\sigma_0\,.  
		 \end{split}\label{Eq:Hwire}
		\end{equation}
		Here, $\tau_k$ and $\sigma_k$ are Pauli matrices in particle-hole and spin 
		space, respectively, and the Nambu basis spinor is given by 
		$(d_\uparrow^\dagger(y),\,d_\downarrow^\dagger(y),\,d_\downarrow(y),
		\,-d_\uparrow(y))$. The parameter $m^*$ is the effective mass of the 
		electrons in the wire, $\alpha_R$ is the Rashba spin-orbit coupling 
		strength, $E_z$ the Zeeman energy due to the perpendicular magnetic field 
		$B_z$, and $\Delta$ the proximity induced s-wave superconducting gap, 
		which we choose to be real. The operator $d^\dagger_j$ creates an electron 
		in the $j$-th eigenstate of $H_{\rm wire}$ in the absence of 
		superconductivity. We treat the charging term 
		\begin{equation}
				H_{\rm ch}
				=\sum_j \left[-eV_g+\frac{E_c}{2}\sum_{i\neq j} 
				 d_i^\dagger d_i\right]d_j^\dagger d_j\ ,
				\label{Eq:charging}
		\end{equation}
		in the Hartree approximation, which yields $E_c(N_0-1)-eV_g$ for the 
		expectation value of the expression in brackets in case of a hole-like 
		co-tunneling process, and $E_cN_0-eV_g$ for	an electron-like process. 
		Here, $E_c$ is the charging energy needed to add an electron to the dot,
		which is proportional to the inverse of the wire length, and $V_g$ is the
		gate voltage. Coupling between dot and leads is described by the 
		tunneling Hamiltonian
		\begin{equation}
			H_{\rm tun}	=\sum_{j\sigma\alpha} t_{\alpha j \sigma} 
			 c_\sigma^\dagger(y_\alpha) d_j + {\rm h.c.}\,, \label{Eq:tunneling}
		\end{equation}
		where the couplings $t_{\alpha j \sigma} 
		= t_0\int{\rm d} y\,\Psi_{\alpha,\sigma}(y)\,\varphi_j(y) $ are  
		approximated as  the overlap integral between a decaying wave 
		$\Psi_{\alpha,\sigma}$ from lead $\alpha$ and the 
		eigenfunction $\varphi_j$ of the Hamiltonian $H_{\rm wire}$ for $\Delta=0$ 
		(see appendix). This approximation is relaxed 
		later where we use a microscopic model to compute the couplings.  
		In the topological regime, the weight of wave function $\Psi_{\alpha\up}$ 
		dominates over $\Psi_{\alpha\down}$, such that the tunneling barrier effectively
		filters one spin direction \citep{Vuik.2019}. In the first part, we 
		therefore focus on the 
		transmission amplitude for spin-up electrons 
		$|T_{\up\up}|$. From the Hamiltonians Eq.~\eqref{Eq:Hwire}, 
		\eqref{Eq:charging}, \eqref{Eq:tunneling} we determine the energies 
		${\varepsilon}_{j}^{e(h)}(N_0,\{n_i\})$ and couplings 
		${\lambda}^{e(h)}_{\alpha j \sigma}(N_0,\{n_i\})$ of lead $\alpha$ by 
		numerically solving the BdG equation, combined with analytical arguments 
		for the spatial parity of wave functions (see appendix). We take into 
		account the particle-hole redundancy in the solutions of the BdG equation 
		by including only the eigenstates with positive energy \citep{Li.2020}. 
		The quantum dot contains an integer number $N_0$ of electrons. On 
		the other hand, the particle number in the wire $N_w$ is fractional in 
		general. We therefore describe the proximity effect using a mean-field 
		superconductivity term in the wire Hamiltonian, but distinguish $N_0$ from 
		$N_w$ \citep{Drukier.2018}. 
		\begin{figure}[t!]
				\centering
				\includegraphics[width=8.6cm]{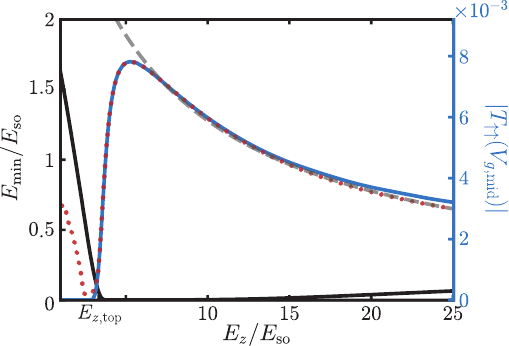} 
				\caption{\label{Fig:AmplitudeConstOPCouplingsAndEnergies} 
						Transmission amplitude $|T_{\up\up}(V_{g,\rm mid})|$ (blue 
						y-axis) for transmission through the lowest effective level only 
						(dotted, red), and $j_{\rm max} = 200$ effective levels (solid, 
						blue), together with the lowest BdG energy (black y-axis, solid 
						black line) of the wire Hamiltonian as a function of the magnetic 
						field. Here $V_{g,\rm mid}$ is the center between amplitude 
						resonances corresponding to a particle number $N_w=35$. 
						The dashed gray line shows the decay of the amplitude according to
						Eq.~\eqref{Eq:RedAmpConstDelta}, where the constant factor is obtained
						from a fit.  
						We use $\Delta =2\,E_{\rm so}$ and $L=32.5\,l_{\rm so}$.
						}  
		\end{figure}	
		The chemical potential $\mu$ is self-consistently 
		determined such that the expectation value of the number of particles in the 
		wire is given by $N_w$, but we take into account the total number of 
		particles in the dot $N_0$ when we determine the charging contribution to 
		the effective single particle energies. 
		When varying the gate voltage one observes a conductance resonance whenever 
		a level crosses the Fermi level. We increase $N_0$ by one after each such 
		resonance. However, it is assumed that only a charge $\Delta N_w \ll 1$ is added 
		to the wire. 
	
	\begin{figure}[ t! ]
			\centering
			\includegraphics[width=8.6cm]{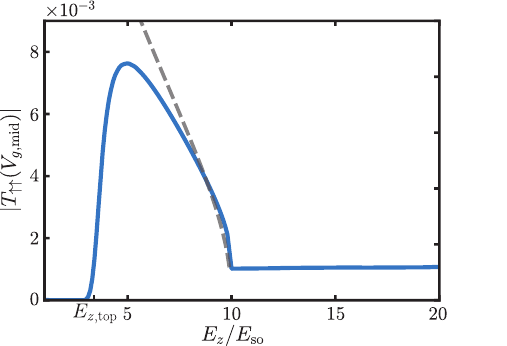}  
			\caption{\label{Fig:BlankWireAmplitudeofN}Transmission amplitude 
					$|T_{\up\up}(V_{g,\rm mid})|$ for $j_{\rm max} = 200$ effective levels 
					as a function of the magnetic field. 
					Here, $V_{g,\rm mid}$  is the center between amplitude resonances 
					corresponding to a particle number $N_w=35$. 
					We use a wire length of $L=32.5\,l_{\rm so}$ and a field dependent 
					induced gap Eq.~\eqref{Eq:SCOP} where  $\Delta(4.5\,E_{\rm so})
					=2\,E_{\rm so}$ and $E_{z,c} = 10\,E_{\rm so}$ (solid, blue). The 
					dashed gray line shows the decay of the amplitude according to 
					Eq.~\eqref{Eq:TWithDeltaOfEz}, where the constant factor is obtained
					from a fit.}
		\end{figure}

\subsection{Parameters}
		For the numerical calculations, we use	a spin orbit coupling 
		strength of $\hbar\alpha_R = 0.2\,{\rm eV\AA}$ and an effective mass  
		$m^* = 0.02\,m_e$, which are typical for semiconductor 
		structures such as InAs \citep{Mourik.2012,Lutchyn.2018}. From these 
		parameters we obtain the characteristic energy and length scales $E_{\rm so} 
		= \alpha_R^2 m^*/2= 0.05\,{\rm meV}$ and $l_{\rm so} =  \hbar/(\alpha_R m^*) 
		= 0.19 \rm \upmu m$, respectively. We discretize the wire Hamiltonian 
		Eq.~\eqref{Eq:Hwire} to $N$ lattice sites with lattice constant
		$a=L/N = 0.026\,l_{\rm so}$, where $L$ denotes the wire length.  
		We assume that each electron that is added to the dot contributes a charge 
		$\Delta N_w = 1/20$ to the wire. We use a charging energy $E_c 
		= 8\,E_{\rm so}\, (32.5\,l_{\rm so})/L$. For computing the average of the 
		scattering matrix Eq.~\eqref{Eq:Weid} we use finite temperature 
		$T= 34\,\rm mK$ \citep{Whiticar.2020} corresponding to $\beta=18\,E_{\rm so}$.  
				
		\begin{figure}[t!] 
				\centering
				\includegraphics[width=8.6cm]{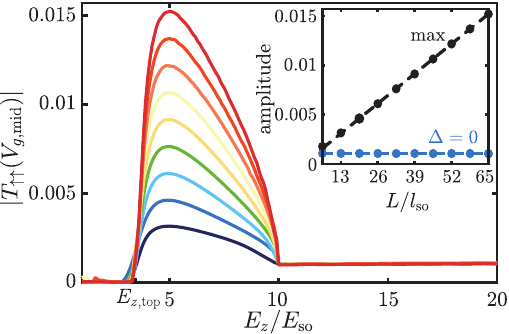}  
				\caption{\label{Fig:AmplitudeofN}Transmission amplitude 
						$|T_{\up\up}(V_{g,\rm mid})|$ as a function of the magnetic 
						field for different wire lengths $L=13\,l_{\rm so},19.5\,
						l_{\rm so}, 26\,l_{\rm so}, 32.5\,l_{\rm so}, 39\,l_{\rm so}, 
						45.5\,l_{\rm so}, 52\,l_{\rm so}, 58.5\,l_{\rm so}$ and 
						$65\,l_{\rm so}$. Here, $V_{g,\rm mid}$ is the center between 
						amplitude resonances for a particle number $N_w
						=35 L/(32.5\,l_{\rm so})$. 
						We assume a Zeeman field dependent gap parameter 
						Eq.~\eqref{Eq:SCOP} with $\Delta(4.5\,E_{\rm so})=2\,E_{\rm so}$ 
						and a critical field $E_{z,c} = 10 E_{\rm so}$. The inset shows 
						the value at the maximum of the amplitude in the topological 
						region (black circles) and the value of the amplitude in the 
						normal-conducting region (blue circles) as a function of the wire 
						length $L$.   
						}
		\end{figure}			
				
\section{Magnetic field dependence of the transmission amplitude}
\subsection{Magnetic field independent induced gap}
		We consider the transmission amplitude $|T_{\up\up}(V_{g,\rm mid})|$ as a 
		function of Zeeman energy $E_z$, computed at a gate voltage $V_{g,\rm mid}$ 
		in the middle between the two conductance resonances for a fixed particle 
		number $N_w=  ( L/l_{\rm so})(14 / 13)$, such that the particle	density is 
		the same for all wire lengths.  
		
		We first consider a magnetic field independent proximity gap $\Delta 
		= 2\,E_{\rm so}$.   
		By increasing $E_z$, the transition to 
		the topological phase takes place, in which an eigenstate close to zero 
		energy is formed,  separated from the second level by the topological gap 
		(see Fig.~\ref{Fig:AmplitudeConstOPCouplingsAndEnergies}). The transmission 
		amplitude strongly increases when entering 
		the topological phase at $E_{z,\rm top}$, reaches a peak value, and then 
		decreases. 
		In the topological regime the tunneling matrix element for a Majorana wave
		function is $\propto 1/\sqrt{\xi}$, where the correlation length 
		\begin{equation}
			\xi = \frac{\hbar v_F}{\Delta_{p,\rm ind}} \label{Eq:xi}
		\end{equation} 
		with $v_F = \hbar k_F(1/m^* - \alpha_{\rm R}^2 (E_z^2+\hbar^2
		\alpha_{\rm R}^2k_F^2)^{-1/2})$ is determined by the induced effective 
		p-wave gap at the Fermi points in the hybrid wire  
		\citep{Kitaev.2001,Lutchyn.2010,Potter.2011}
		\begin{equation}
			\Delta_{p,\rm ind} 
			 = \frac{\hbar k_F\alpha_R\Delta}{\sqrt{E_z^2+\alpha_R^2\hbar^2k_F^2}} 
			 \ . \label{Eq:DeltaInd}
		\end{equation}
		With this, we obtain the Zeeman field dependence of the transmission amplitude as 
		\begin{align}
			|T_{\up\up}| \sim \frac{m^*\alpha_{\rm R}\Delta}{\hbar}
			\frac{1}{\sqrt{E_z^2+\alpha_{\rm R}^2\hbar^2k_F^2}-\alpha_{\rm R}^2 m^*}\;,  
			\label{Eq:RedAmpConstDelta}
		\end{align}
		proportional to the inverse field strength for large $E_z$ (dashed gray line in 
		Fig.~\ref{Fig:AmplitudeConstOPCouplingsAndEnergies}, in very good agreement with 
		the numerical result taking a single level into account). When comparing 
		the result for transmission through $j_{\rm max}=200$ levels  (solid blue 
		line) with that for a single level 
		(Fig.~\ref{Fig:AmplitudeConstOPCouplingsAndEnergies}, dotted red line) 
		it becomes apparent 
		that the amplitude at the beginning of the topological range is mostly 
		determined by the lowest level, i.e.~the MZMs. 
		For very large Zeeman energy, the Majorana modes are split more strongly, and there is a small
		correction due to taking into account many higher levels.  
		In the trivial regime for $E_z<E_{z,\rm top}$ however, where the spacing between 
		the lowest energy Bogolubons is small, many levels contribute to the
		transmission amplitude, and interfere destructively.

	\begin{figure}[t!]	
				\centering
				\includegraphics[width=8.6cm]{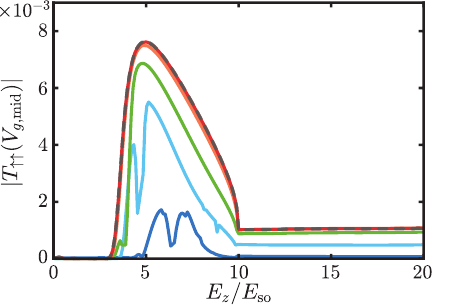}
				\caption{\label{Fig:Disorder}Transmission amplitude 
						$|T_{\up\up}(V_{g,\rm mid})|$ as a function of the magnetic 
						field for various strength of on-site disorder. We use a
						wire length of $L=32.5\,l_{\rm so}$ and compute the 
						amplitude between resonances corresponding to particle 
						number $N_w=35$. The gray dashed line is for reference 
						without disorder. The colored lines (from top to bottom) 
						are numerically computed for Gaussian disorder with 
						standard deviation $W=0.1\,E_{\rm so} \ll W_{\rm m}$, 
						$W=1\,E_{\rm so}$, $W=5\,E_{\rm so}$, $W=10\,E_{\rm so} 
						\sim W_{\rm m}$, and $W=20\,E_{\rm so}$. }
		\end{figure} 	
		
		\begin{figure*}[t!]	
			\centering
			\includegraphics[width=17.2cm]{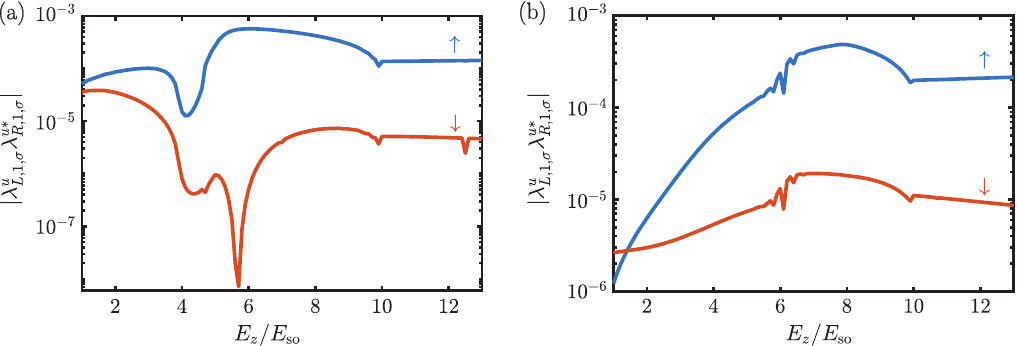}
			\caption{\label{Fig:CouplingsSpinResolved}Dot-lead couplings for 
					both spin directions using a wire of length 
					$L=45.5\,l_{\rm so}$. We use the Zeeman field 
					dependent induced gap $\Delta(E_z)$  Eq.~\eqref{Eq:SCOP} 
					with a critical field $E_{z,c}=10\,E_{\rm so}$.
					(a) Couplings to the lowest dot level using the steep 
					potential and particle number $N_w=47L/(39\,l_{\rm so})$
					in the wire. (b) Couplings to the  lowest dot level 
					using the smooth potential for $N_w=53L/(39\,l_{\rm so})$. }
		\end{figure*}	
				
 \subsection{Magnetic field dependent induced gap}
		For a thin superconductor subject to a parallel field,  we describe 
		the suppression of the induced s-wave  superconducting gap by the magnetic field via 
		\citep{Tinkham.2004}
		\begin{equation}
			\Delta(E_z) = \Delta(0) \left[ 1-\left( \frac{E_z}{E_{z,c}}\right)^2 
			 \right]^{1/2}\;,  
			\label{Eq:SCOP}
		\end{equation}
		where $E_{z,c}$ is the critical Zeeman energy at which superconductivity   
		is destroyed. Entering the topological region at $E_{z,\rm top}$ is again 
		accompanied by an increase in transmission amplitude (see 
		Fig.~\ref{Fig:BlankWireAmplitudeofN}).  
		Further within the topological regime, the proximity gap $\Delta$  is reduced,
		and the correlation length $\xi$ 
		$\propto 1/|\Delta_{p,\rm ind}|$  
		increases, i.e. the Majorana wave function delocalizes.  
		Therefore, the amplitude drops to the normal-conducting value  over a 
		relatively narrow range of magnetic field values.	Using Eq.~(\ref{Eq:SCOP}) 
		in Eq.~\eqref{Eq:DeltaInd}, we find an amplitude dependence
		\begin{equation}
			|T_{\up\up }|  \sim \frac{m^*\alpha_R \Delta(0)}{\hbar}
			\frac{\sqrt{1-\left(  {E_z}/{E_{z,c}}\right)^2}}
			 {\sqrt{E_z^2+\alpha_R^2\hbar^2k_F^2}-\alpha_{\rm R}^2m^*} \ .
			\label{Eq:TWithDeltaOfEz}
		\end{equation} 
		This dependence is depicted by the dashed gray line and fits well in 
		the region where the amplitude decays to the normal-conducting value (see 
		Fig.~\ref{Fig:BlankWireAmplitudeofN}). For $E_z> E_{z,c}$, the wire is 
		normal-conducting, and the amplitude is approximately constant. These 
		results for the amplitude are in agreement with the  recent 
		experiment \citep{Whiticar.2020}.

\subsection{Wire length dependence}
		The non-locality of MZMs is expected to have a profound consequence 
		when considering wires of varying lengths. In the inset of 
		Fig.~\ref{Fig:AmplitudeofN}, the value of the amplitudes 
		at the maximum and in the normal-conducting region are depicted as a 
		function of the wire length $L$. From our scattering matrix 
		analysis using a charging energy that is proportional to the inverse of the
		wire length, we find that the transmission amplitude is indeed 
		proportional to the wire length in the topological region, while it 
		is independent of the wire length in the normal-conducting range
		(see Fig.~\ref{Fig:AmplitudeofN}).

\section{Disorder in the wire}		
		The proposed experiment for establishing the wire length dependence 
		of the transmission amplitude in the presence or absence of MZMs 
		requires the comparison of different wires. Since these wires may 
		differ from each other in terms of their detailed composition, we 
		study how robust our results for the transmission amplitude are in 
		the presence of on-site disorder. We use a Gaussian	disorder 
		distribution with zero mean and standard deviation $W$. Disorder 
		is strong when the elastic scattering rate $\hbar/\tau$ from the 
		impurities is on the order of the induced effective gap 
		$\Delta_{p,\rm ind}$ in the wire \citep{Lynton.1957,Pippard.1957,
		Anderson.1959,Buchholtz.1981,Hirschfeld.1986,SchmittRink.1986,
		Maki.1999,Zocher.2013a}. We define a critical disorder strength 
		$W_{\rm m}$ such that the effect of disorder on the amplitude is 
		negligible for $W\ll W_{\rm m}$. For $W \approx W_{\rm m}$
		disorder has noticeable effects on the amplitude and for 
		$W\gg W_{\rm m}$ pair breaking sets in and destroys the	
		superconducting properties and the amplitude vanishes. Using 
		Fermi's golden rule, we estimate the elastic scattering rate for 
		the case of a scatterer at each lattice site as
		\begin{equation}
			\frac{\hbar/\tau}{E_{\rm so}} 
			= \left(\frac{W}{E_{\rm so}}\right)^2 \frac{a}{l_{\rm so}} 
			 \frac{1}{k_F l_{\rm so}} \ .
		\end{equation}
		The induced gap at the Fermi momentum $k_Fl_{\rm so} = 
		 (2+{\mu}/{E_{\rm so}}+\left[({E_z}/{E_{\rm so}})^2 + 4 
		+ 4 {\mu}/{E_{\rm so}}\right]^{1/2} )^{1/2} $ is given by
		\begin{equation}
			\frac{\Delta_{p, \rm ind}}{E_{\rm so}}  
			= 2\frac{\Delta}{E_{\rm so}} 
			\frac{k_Fl_{\rm so}}{\sqrt{(E_z/E_{\rm so})^2+4(k_Fl_{\rm so})^2}}\ .
		\end{equation}
		We define $W_{\rm m}$ such that $\hbar/\tau=\Delta_{p,\rm ind}$ for 
		$W=W_{\rm m}$, i.e.
		\begin{equation}
			\frac{W_{\rm m}}{E_{\rm so}} 
			 = \sqrt{2\frac{l_{\rm so}}{a}\frac{\Delta}{E_{\rm so}}}
			\frac{k_Fl_{\rm so}}{\left[ (E_z/E_{\rm so})^2 
			 + 4 (k_Fl_{\rm so})^2 \right]^{1/4}}\ .
		\end{equation}
		Numerical results of the amplitude for various disorder strengths 
		are depicted in Fig.~\ref{Fig:Disorder}. When the disorder 
		strength is smaller but of the order of $W_{\rm m}$, the 
		transmission amplitude is reduced at its maximum. This reduction 
		is however much smaller that the peak height such that the proposed 
		experiment is robust against disorder $W < W_{\rm m}$. When using 
		a disorder strength close to or larger than $W_{\rm m}$, the 
		amplitude is significantly reduced.

		\begin{figure*}[t!]	
			\centering
			\includegraphics[width=17.2cm]{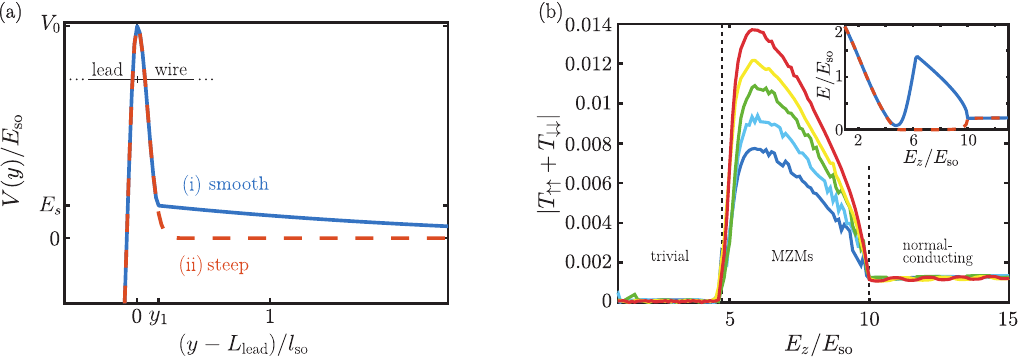}   
			\caption{\label{Fig:AmplitudeMicroscopCouplings} 
					(a) Barrier potential used to compute the lead-wire couplings in 
					the microscopic model. The leads of length $L_{\rm lead}$ are 
					normal-conducting and without spin-orbit coupling. At position $y_1$ 
					and energy $E_s$ the narrow Gaussian peak transitions continuously into
					the wide peak in the case of the smooth potential. The height of the peak 
					is given by $V_0=65\,E_{\rm so}$.
					(b) Numerical results for the amplitude $|T_{\up\up}(V_{g,\rm mid})
					+T_{\down\down}(V_{g,\rm mid})|$ with microscopic couplings as a function
					of the Zeeman field using a steep confinement 
					potential and ground state particle number $N_w=47L/(39\,l_{\rm so})$. 
					We consider wires of length $L=32.5\,l_{\rm so}, 
					39\,l_{\rm so}, 45.5\,l_{\rm so}, 52\,l_{\rm so}$, and 
					$58.5\,l_{\rm so}$.
					The
					results are in good agreement with Fig.~4 where we used the more 
					paradigmatic model for the couplings. 
					}
		 \end{figure*}

\section{Microscopic model for couplings}		
		In the first part, we assumed that the couplings 
		between lead and dot are determined by the dot wave functions at
		the ends of the wire. To validate this assumption, we consider a 
		tight-binding model of leads and wire which are separated by tunnel
		barriers of shape	$V_{\sigma_i,V_0}(y) 
		= V_0\exp(-y^2/(2\sigma_i^2))$.   
		The potential at 
		the left lead is given by
		\begin{align}
			V(y) &=\begin{cases}  
				V_{\sigma_1,V_0+V_{\rm lead}}(y-L_{\rm lead})-V_{\rm lead} 
					&  y\leq L_{\rm lead}\\
				V_{\sigma_1,V_0}(y-L_{\rm lead})                               
					&  L_{\rm lead} < y < y_1\\	
				V_{\sigma_2,V_0}(y-y_1+y_2-L_{\rm lead})                        
					&  y \geq y_1\\
			\end{cases}  \notag\\
			y_j  &=\sqrt{2\sigma_j^2\ln(V_0/E_s)} \ , \label{Eq:ABSPot}
		\end{align}
		where $L$ and $L_{\rm lead}$ are the length of wire and leads, 
		respectively. The potential in the leads is lowered by an offset 
		$V_{\rm lead} = 100\,E_{\rm so}$, such that both spin directions
		are present at the Fermi level. The leads are normal-conducting and
		without spin-orbit coupling.  In the wire the potential 
		consists of two parts with standard deviations $\sigma_1$ and 
		$\sigma_2$ which are matched continuously at $(y_1,E_s)$ by 
		shifting the second peak by $y_1-y_2$. 
		
		We define microscopic couplings as matrix elements
		$
		 \lambda_{\alpha i \sigma}^u  
			= \bra{\bm{\Phi}_{\alpha\sigma}^u} H \ket{\bm{\Psi}_i}  
		 \text{ and }
		 \lambda_{\alpha i \sigma}^v  
			= \bra{\bm{\Phi}_{\alpha\sigma}^v} H \ket{\bm{\Psi}_i}\ 
		$
		of the combined Hamiltonian of wire and leads.
		Here, $\bm{\Psi}_i$ is the $i$-th BdG level in the wire, and due to
		the particle-hole symmetry and the absence of superconductivity
		in the leads we can write
		$\bm{\Phi}_{\alpha\sigma}^u = (\varphi_{\alpha\up}^{(\varepsilon_F,\sigma)},
		\varphi_{\alpha\down}^{(\varepsilon_F,\sigma)},0,0 )$ and 
		$\bm{\Phi}_{\alpha\sigma}^v = (0,0,
		\varphi_{\alpha\down}^{(\varepsilon_F,\sigma)*},
		-\varphi_{\alpha\up}^{(\varepsilon_F,\sigma)*})$ where 
		$ ( \varphi_{\alpha\up}^{(\varepsilon_F,\sigma)},
		\varphi_{\alpha\down}^{(\varepsilon_F,\sigma)}  )$ is the wave 
		function localized in lead $\alpha$ with spin $\sigma$ that is closest to 
		the Fermi level. To numerically obtain the 
		wave function localized in one region, we fix the potential 
		at height $V_0$ in all the other regions.  
		Fig.~\ref{Fig:CouplingsSpinResolved} depicts the 
		couplings between lead and first dot level for spin-$\up$ and 
		spin-$\down$ electrons. For both types of confinement, the 
		couplings of spin-$\up$ electrons are dominant. It is therfore justified to only
		consider $|T_{\up\up}|$ in the more paradigmatic model in 
		the first part.
		Due to the effective time-reversal symmetry $T = \sigma_z K$ with $K$ denoting 
		complex conjugation, $T^2 = +1$, and $[H,T] = 0$, the transmission amplitudes 
		$T_{\up\up}$ and $T_{\down\down}$ have both the same phase (modulo $\pi$), such 
		that the magnitude of the interference term in Eq.~(1) is given by  
		$|T_{\up\up}+T_{\down\down}|$. We compute this magnitude for various wire lengths.\\
		\begin{figure*}[t!]	
			\centering
			\includegraphics[width=17.2cm]{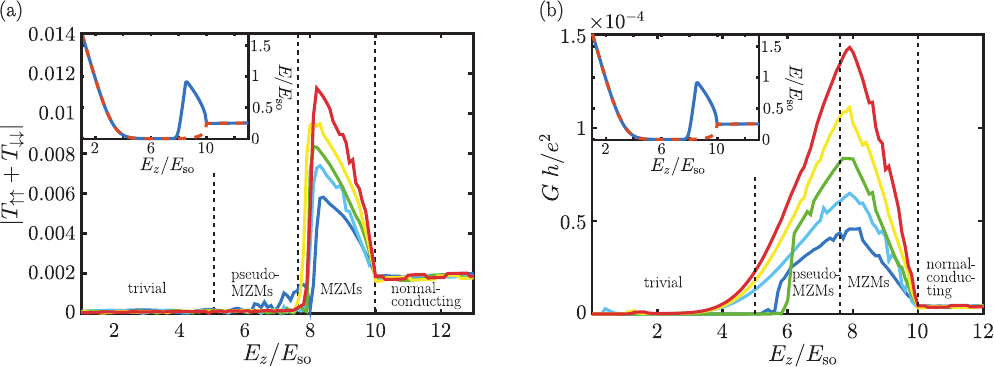}
			\caption{\label{Fig:AmplitudeSteep}
					(a) Numerical results for the transmission amplitude 
					$|T_{\up\up}(V_{g,\rm mid})+ T_{\down\down}(V_{g,\rm mid})|$  using 
					tunnel couplings obtained for the smooth barrier potential  
					as a function of the Zeeman energy.  
					(b) Conductance through the wire without an interferometer in the case of
					a smooth potential for $N_w = 53 L/(39\,l_{\rm so})$. A comparison with (a)
					shows that the interferometer is crucial to distinguish
					MZMs from pseudo-MZMs.
					We consider wires of length $L=32.5\,l_{\rm so}, 
					39\,l_{\rm so}, 45.5\,l_{\rm so}, 52\,l_{\rm so}$, and 
					$58.5\,l_{\rm so}$. We use a Zeeman field dependent
					induced gap Eq.~\eqref{Eq:SCOP} with a critical field $E_{z,c}
					=10\,E_{\rm so}$ and a particle number $N_w = 53 L/(39\,l_{\rm so})$.
					The insets depicts the lowest two energy eigenvalues of the wire 
					Hamiltonian for $L= 45.5\,l_{\rm so}$. 
					}
		\end{figure*} 
    \indent We distinguish two types of barrier potentials (i) only a narrow 
		Gaussian peak and (ii) a narrow Gaussian peak together with a potential decaying 
		smoothly into the wire (see Fig.~\ref{Fig:AmplitudeMicroscopCouplings}(a)).  
		Parameters for (i) the steep potential are given by $\sigma_1=\sigma_2=0.1
		\,l_{\rm so}$, $E_s = V_0$ and $V_0=65\,E_{\rm so}$, and for (ii) 
		the smooth confinement $\sigma_1=0.1\,l_{\rm so}$, 
		$\sigma_2=6\,l_{\rm so}$, $E_s = 10\,E_{\rm so}$ and $V_0=65
		\,E_{\rm so}$.  
		In case (i) there are zero-energy states only 
		in the topological region, which are the MZMs 
		(see inset of Fig.~\ref{Fig:AmplitudeMicroscopCouplings}(b)). 
		In case (ii), the Fourier 
		decomposition of the smooth potential does not contain large momenta, so 
		that in the trivial region each of the two bands contributes a pair of 
		MZMs, which however are not coupled among each other by the potential 
		\citep{Prada.2012,Kells.2012, Vuik.2019}. Therefore, in addition to the 
		MZMs in the topological region, two quasi-degenerate, quasi-zero energy Andreev
		bound states (also called pseudo-MZMs) occur in the trivial region for 
		$5\,E_{\rm so}<E_z<7.6\,E_{\rm so}$ (see inset of 
		Fig.~\ref{Fig:AmplitudeMicroscopCouplings}(b)).  
		Since they are nearly degenerate, there are two ground states
		with equal Boltzmann weight in the thermal average. 
		For even $N_0$ the 
		degenerate ground states for $\mathcal{E}_1=\mathcal{E}_2=0$ are 
		states where either all $N_0$ electrons are in the condensate or 
		$N_0-2$ electrons form the condensate and both pseudo-MZMs are
		occupied. In the case of odd $N_0$ there are $N_0-1$ electrons in the 
		condensate and either the first or the second pseudo-Majorana level is	occupied. 
		In both cases the thermally averaged amplitude is proportional to 
		$\sum_{j=1}^2(\lambda_{L,j,\up}^u\lambda_{R,j,\up}^{u*}
		+\lambda_{L,j,\up}^v\lambda_{R,j,\up}^{v*}) \approx 0$. The 
		anti-unitary reflection symmetry  
		$\tilde{\Pi} \varphi_j(y)=K\varphi_j(L-y)$ (where $\varphi_j$ are 
		eigenfunctions of $H_{\rm wire}$) ensures that both terms are real and	
		${\rm sgn}(\lambda_{L,j,\up}^u\lambda_{R,j,\up}^{u*})
		=-{\rm sgn}(\lambda_{L,j,\up}^v\lambda_{R,j,\up}^{v*})$ 
		\citep{Drukier.2018}. Due to the Majorana condition for zero energy states  
		$|u_{j\sigma}|=|v_{j\sigma}|$ the terms 
		cancel each other. Hence, the ground state degeneracy gives rise to a vanishing amplitude 
		upon  thermal averaging \citep{Hell.2018}: Forming a Cooper pair or 
		occupying the two zero-energy pseudo-MZMs requires the same energy, but 
		yields contributions with opposite signs and equal magnitude to the 
		transmission amplitude.   
		For a  wire of finite length, 
		the pseudo Majorana modes do not lie exactly at zero energy  
		and a finite amplitude is observed. This is the case for the smallest wire 
		length in Fig.~\ref{Fig:AmplitudeSteep}(a). As long as 
		the two levels are nearly degenerate and nearly at zero energy, the 
		amplitude is well below the $\Delta=0$ value and no pronounced 
		maximum is formed. In addition, the amplitude is not proportional to 
		$L$ in the pseudo-MZM regime.	
		
		In comparison with the more paradigmatic model
		considered before, we find that for a steep potential (i) all qualitative features of 
		the amplitude remain unchanged (Fig.~\ref{Fig:AmplitudeMicroscopCouplings}(b)). 
		Importantly, the suppression of the transmission amplitude in the trivial regime
		occurs even  when pseudo-MZMs are present (see	Fig.~\ref{Fig:AmplitudeSteep}(b)).  \vspace{-0.15cm}

\section{Comparison between interferometer setup and direct conductance measurement} 
		In this section, we compare signatures from the interferometer setup
		(Fig.~\ref{Fig:SetupAAA}), with an easier to implement direct 
		conductance measurement through the dot, without interferometer. 
		In the calculation of the transmission amplitude 
		through the dot (interferometer case) or the transmission probability (direct conductance), 
		the main difference is how the thermal average is 
		performed. For the calculation of the amplitude of conductance oscillations 
		through the interferometer, the thermal average is performed over the complex 
		transmission amplitude (see Eq.~\eqref{Eq:Weid}), so that the 
		transmission phase contributes to such an average. In the case of a direct conductance 
		measurement, the squared absolute value of the transmission amplitude is 
		averaged, and the phase information does not contribute. Fig.~\ref{Fig:AmplitudeSteep}(b) 
		depicts the direct conductance through the dot for a smooth confinement 
		potential, analogous to Fig.~\ref{Fig:AmplitudeSteep}(a). 
		For MZMs we also find a maximum at the beginning of the topological region, 
		whose height scales with the wire length. The crucial difference is 
		that the conductance is not suppressed for pseudo-MZMs and thus this maximum 
		is not a unique signature for the presence of MZMs.	\vspace{-0.15cm}

\section{Connection to experiment}
		\begin{figure}[t!] 
				\centering
				\includegraphics[width=8.6cm]{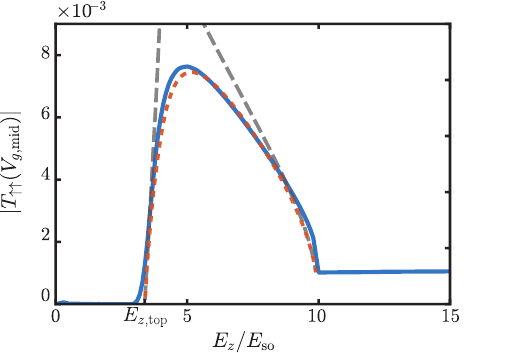}  
				\caption{\label{Fig:AmplitudeApproximation} 
					Transmission amplitude 
					$|T_{\up\up}(V_{g,\rm mid})|$ for $j_{\rm max} = 200$ effective levels 
					(solid, blue)
					as a function of the magnetic field with parameters as in 
					Fig.~\ref{Fig:BlankWireAmplitudeofN}. 
					The dashed gray lines depict the $\xi_s$ approximation at the beginning of the
					topological regime and the $\xi$ approximation in the region where superconductivity
					is destroyed by the magnetic field. The dotted red line depicts the approximation
					Eq.~\eqref{Eq:aproxTmax}, where both correlation lengths are taken into account.
					The maximum of the transmission amplitude arises due to the interplay of both terms.
					We use the same proportionality  constant for all three approximations obtained by a
					fit.
						}
		\end{figure}
		
		In a recent experiment by Whiticar et al.~\citep{Whiticar.2020},  the transmission 
		amplitude through a Coulomb blockaded Majorana wire was measured as a function of 
		the Zeeman field. The experimental transmission amplitude  shows a rapid growth upon entering 
		the topological regime,   followed by  a pronounced  maximum.  Here, we discuss in detail how 
		these features are explained by the localization properties of MZMs, which determine the 
		transmission amplitude in the topological regime. 
		
		In the case of sufficiently long wires, in which the Majorana wave 
		functions of opposite wire ends have negligible overlap, an analytical 
		solution for the MZM wave functions can be found (see Appendix B). 
		Moreover, since the transmission amplitude in the 
		topological region is determined almost exclusively by transport 
		through MZMs,   the transmission amplitude can directly be obtained
		from the Majorana wave functions.
		In Section III, we  discussed that for large Zeeman fields, deep in the topological regime,  
		the spatial decay of MZMs is characterized by 
		the p-wave localization length 
		$\xi = \hbar v_F/\Delta_{p,\rm ind}$ Eq.~\eqref{Eq:xi}. However, 
		from the full analytic solution is is apparent that  there is a second localization length 
		%
		\begin{eqnarray}
				\xi_s  
				&= & \left( -\xi^{-1} + \sqrt{\xi^{-2} - \frac{\mu^2+\Delta^2-E_z^2}{(\xi^{-2}+k_F^2) E^2_{\rm so}}} \right)^{-1}   
				\nonumber \\
				&\propto & {1 \over E_z-\sqrt{\Delta^2+\mu^2} }\ ,
		\end{eqnarray}
		%
		which describes the localization properties of MZMs for Zeeman fields $E_z \gtrsim E_{z, \rm top}$ 
		close to the topological phase transition.  We can  approximate the envelope of the Majorana wave function by 
		a sum of two exponentially decaying terms (for details see Appendix B)
		\begin{align}
					\chi_{L,\uparrow} \approx\frac{1}{\sqrt{\xi+\xi_s}}\left(
						\e^{-y/\xi} + \e^{-y/\xi_s}\right)\ . \label{Eq:aproxMWF}
		\end{align}
		The corresponding Majorana wave function at the right end is then given by 
		$\chi_{R,\uparrow}(y)\propto \chi_{L,\uparrow}(L-y)$. This yields for
		the transmission amplitude 
		\begin{align}
			|T_{\uparrow\uparrow}(V_{g,\rm mid})| \propto \frac{1}{\xi+\xi_s} \ . \label{Eq:aproxTmax}
		\end{align}

		 A comparison shows that the approximated transmission amplitude 
		(dotted, red line in Fig.~\ref{Fig:AmplitudeApproximation}) is in very good 
		agreement with the numerical results for transmission through 200 
		levels (solid, blue line in Fig.~\ref{Fig:AmplitudeApproximation}). Thus, the 
		competition of the two correlation lengths $\xi$ and $\xi_s$ 
		explains the occurrence of the maximum in the transmission amplitude. 
		
		In addition, 
		in Fig.~\ref{Fig:AmplitudeApproximation}, 
		we compare the numerically obtained 
		transmission amplitude 
		to approximations taking into account the larger of the two localization lengths:
		The behavior of the transmission amplitude at the beginning 
		of the topological region can be understood by the  localization length 
		$\xi_s$ alone, i.e. $|T_{\uparrow\uparrow}|\propto 1/\xi_s$ (dashed gray line in the beginning
		of the topological regime). On the other hand, the behavior near the transition into the normal-conducting 
		region, is due to the p-wave localization length $\xi $, i.e.~$|T_{\uparrow\uparrow}|\propto 1/\xi$
		(dashed gray line at the end of the topological regime). 
		 The maximum occurs where the magnitude 
		of the localization lengths is roughly comparable.

		The picture described above allows to explain the magnetic field dependence of transmission amplitude
		found by Whiticar et al.~\citep{Whiticar.2020}. 
		In the experiment, the transmission amplitude depends only weakly on the magnetic field in the region 
		of small Zeeman fields, as predicted for the trivial phase. Above a device-specific value of the magnetic 
		field, a rapid increase of the transmission amplitude is observed, which can be explained by the magnetic 
		field dependence of $1/\xi_s$ at the beginning of the topological phase. Due to the divergence of $\xi_s$ at the phase transition 
		$E_z=E_{z,\rm top}$, the transmission amplitude increases linearly  $|T_{\uparrow\uparrow}|\propto E_z-E_{z,\rm top}$ 
		in the topological regime.  For larger Zeeman fields, 
		a well-defined maximum of the amplitude arises in the experiment, which can be understood in terms of the 
		concurrence of both correlation lengths $\xi$ and $\xi_s$. When superconductivity is destroyed by the 
		magnetic field, Whiticar et al.~observe a rapid decline of the transmission amplitude. This decrease 
		can be explained in our model by the divergence of the coherence length $\xi$ due to the vanishing of 
		the induced p-wave gap when approaching the critical magnetic field.

		Since the amplitude of coherent transmission does not exhibit a maximum in the case of pseudo-MZMs, we 
		believe that it is very likely that genuine topological MZMs were observed in the experiment. This is 
		further supported by the observation that together with the appearance of the maximum also the even-odd splitting of the 
		conductance resonances is suppressed. While the behavior of the transmission amplitude in the topological 
		regime can be understood with our one-dimensional model, it is currently not possible to explain the large 
		ratio between the value of the transmission amplitude at the maximum and the value in the normal-conducting 
		regime for Device 2 measured by Whiticar et al. This could be because the amplitude in the experiment is 
		not corrected for the influence of the transmission through the reference arm. On the other hand, it might 
		be necessary to include the influence of orbital effects and several transverse subbands in the theoretical 
		calculations for quantitative agreement between theory and experiment.

		We believe that the experimental results are a promising step towards a proof for the presence of MZMs. Further 
		evidence that MZMs can be consistently observed in these devices would be provided by a systematic   study of wires 
		with different lengths in future experiments.

\section{Conclusion} 
		We have studied coherent transport of electrons through a system 
		hosting MZMs. We find that the Zeeman field and length dependence 
		of the transmission amplitude provide unique signatures of MZMs.
		When considering wires of varying lengths, the 
		non-locality of MZMs yields a stable maximum of the amplitude 
		at the onset of the topological regime, whose height is proportional 
		to the wire length. In contrast, the amplitude is independent of 
		the wire length if no localized MZM is present.  \vspace{-0.15cm}

\begin{acknowledgments}
 We would like to thank C.~Marcus for helpful discussions. 
 This work has been funded by the Deutsche Forschungsgemeinschaft 
 (DFG) under Grant Nos. RO 2247/11-1 and 406116891 within the 
 Research Training Group RTG 2522/1.
\end{acknowledgments}

\renewcommand{\theequation}{A\arabic{equation}}
\setcounter{equation}{0}

\section*{APPENDIX A: Details on scattering matrix formalism}
\subsection*{Truncation of the Hilbert space}
		As described in the main text, we do not explicitly model the 
		superconductor but account for the proximity effect by including 
		the induced superconducting gap directly into the Hamiltonian of 
		the wire. However, we distinguish the particle number in the wire 
		$N_w$ from that in the dot $N_0$ consisting of wire and 
		superconductor.  
		Due to the Coulomb repulsion, simultaneous tunneling of more than one 
		electron or hole is suppressed. We therefore truncate the Hilbert 
		space to states of $N_0$ particles and states of $N_0+1$ electrons for electron-like 
		co-tunneling processes and $N_0-1$ for hole-like co-tunneling,  
		respectively,  but take into account many BdG eigenstates. 
		We denote the occupation number of the $j$-th BdG eigenstate by 
		$n_j$. We introduce states $\ket{N_0,\{n_i\}}$ and 
		$\ket{N_0\pm 1,\{n_i'\}}$ where the former is the initial dot 
		state with $N_0$ electrons and occupation of BdG quasi-particle 
		states $\{n_i\}$. The latter is the intermediate, excited state 
		with $N_0\pm 1$ electrons, and occupation numbers $\{n_i'\}$.  
		As a result of the mean-field  treatment of the 
		interaction in the BCS approach, 
		the theory does not describe a definite 
		particle number $N_0$ in the dot. However, fixed-$N_0$ 
		superconducting systems can even in	case of small $N_0$ be 
		adequately described in the grand-canonical BCS theory by 
		choosing the chemical potential $\mu$ such that the mean 
		particle number $\ex{\hat{N}_0}_\mu$ is given by $N_0$ 
		\citep{Braun.1999}.	 
		We determine the 
		chemical potentials self-consistently for the particle  number  
		$N_w$ in the wire and use the dot particle number $N_0$ for computing
		the charging energy. We note that as the couplings only 
		depend on $N_w$ and since we evaluate the transmission amplitude 
		for a	gate voltage in  between conductance resonances, which are determined by 
		the lowest effective hole-like and electron-like level, the 
		amplitude does not depend on $N_0$ but only on $N_w$. We therefore 
		define the self-consistently determined chemical potential 
		$\mu \equiv \mu({N_0},\{n_i\})$ such that 
		\begin{align}
			\ex{\hat{N}}_{\mu} 
			&\equiv \int{\rm d} y\Bigg[\sum_{j=1}^{2N} |\bm{v}_j(y,\mu)|^2 
			+\sum_{\substack{j,  n_j=1}}\big( -|\bm{v}_j(y,\mu)|^2\notag\\
			&\phantom{-----}
			+|\bm{u}_j(y,\mu)|^2 \big) \Bigg]
			= N_w  \label{Eq:SupN0}\ .
		\end{align}
		Here, due to the particle hole symmetry $P=\tau_y\otimes\sigma_y K$,
		we only need eigenfunctions with non-negative eigenenergies 
		$\mathcal{E}_j(\mu)\geq 0$, which solve the BdG 
		equation 
		\begin{equation}
			H_{\rm wire}(\mu) \begin{pmatrix}{\bm{u}_j(\mu)}\\
			 {\bm{v}_j(\mu)}\end{pmatrix}
			= \mathcal{E}_j(\mu)\begin{pmatrix}{\bm{u}_j(\mu)}
			 \\{\bm{v}_j(\mu)}\end{pmatrix} \ \ .
		\end{equation}
		We rank order the energies and corresponding wave functions such 
		that $\mathcal{E}_1 = \min\{\mathcal{E}_j\}$ and $\mathcal{E}_{j+1}
		>\mathcal{E}_j$. An important exception  to this rule occurs when 
		the wire is in the topological regime $|\mu|<\sqrt{E_z^2-\Delta^2}$, 
		where two Majorana sub-gap states are present in the full BdG 
		spectrum. As the two Majorana wave functions overlap in a wire of 
		finite length $L$, they hybridize to form a finite energy sub-gap 
		BdG state $\mathcal{E}_1$. Increasing the chemical potential, one 
		observes that this energy adiabatically evolves into a negative 
		excitation energy $-|\mathcal{E}_1|$, which corresponds to a change 
		in parity of the ground state. We then need to take the 
		solution with $\mathcal{E}_{-1}=-\mathcal{E}_1$ and $(\bm{u}_{-1},
		\bm{v}_{-1})=(\bm{v}^*_{1},\bm{u}^*_{1})$ as the lowest level, 
		because it corresponds to the odd parity ground state in the 
		topological regime \citep{Drukier.2018}. We denote this solution 
		again by $\mathcal{E}_{1}, (\bm{u}_{1},\bm{v}_{1})$ and use it for 
		the corresponding chemical potentials in the computation of 
		effective couplings and the effective energies. \\ 
		We next express the  wire and tunneling Hamiltonian in terms of 
		BdG  operators $\beta_j$. In order to do so, we first relate  
		annihilation and creation operators $d_{j,\sigma}$, 
		$d_{j,\sigma}^\dagger$ to eigenfunctions $\varphi_{j,\sigma}$ with 
		spin $\sigma$  of $H_{\rm wire}$ for $\Delta =0$ via $d_{j,\sigma} 
		=\sum_{\sigma}\int{\rm d }y\;\varphi^*_{j,\sigma}(y)\,\Psi_\sigma(y).$ 
		Using the expansion of field operators $\Psi_\sigma(y)
		=\sum_j \e^{-i\frac{\phi}{2}}\left[u_{j\sigma}(y,\mu)\beta_j(\mu)
		+v_{j\sigma}^*(y,\mu)\beta_j^\dagger(\mu)\right]$ one finds  
		\begin{align} 
			H_{\rm wire}&=\sum_{\substack{j\\\mathcal{E}_j\geq 0}}
			 \mathcal{E}_j(\mu) \beta_j^\dagger(\mu)\beta_j(\mu) \\
			H_{\rm tun}	&=\sum_{j\sigma\alpha} c_\sigma^\dagger(y_\alpha) 
			\e^{-i\frac{\phi}{2}}\Big[\lambda_{j\alpha\sigma'}^{u}(\mu)\beta_j(\mu) \notag\\
			&\phantom{----} +\lambda_{j\alpha\sigma'}^{v} (\mu)\beta_j^\dagger(\mu)\Big]
			+{\rm h.c.}\; . \label{Seq:HTunMapSc} 
		\intertext{Here the effective couplings are defined by}
			\begin{split}
				\lambda_{j\sigma\alpha}^u(\mu)&=\sum_{i \sigma'}\int{\rm d} y\;
				t_{i\alpha\sigma}\varphi_{i\sigma'}^*(y)\,u_{j\sigma'}(y,\mu)\,, \\
				\lambda_{j\sigma\alpha}^v(\mu)&=\sum_{i \sigma'}\int{\rm d} y\;
				t_{i\alpha\sigma}\varphi_{i\sigma'}^*(y)\,v^*_{j\sigma'}(y,\mu)\;.
		\end{split} \label{SEq:CouplSCDef}
		\end{align}
		In the first part,  the couplings 
		 $t_{\alpha j \sigma} 
		= t_0\int{\rm d} y\,\Psi_{\alpha,\sigma}(y)\,\varphi_j(y) $ are  
		approximated as  the overlap integral between a decaying wave 
		$\Psi_{\alpha,\sigma}$ from lead $\alpha$ and the 
		eigenfunction $\varphi_j$ of the Hamiltonian $H_{\rm wire}$ for $\Delta=0$.
		We take $\Psi_{L,\sigma}\propto \exp(-y/\lambda)$ with $\lambda=0.26\,l_{\rm so}$
		and similar for the right end.
		Since the couplings $t_{j\alpha\sigma}$ are therefore mostly determined by 
		the values of the wave functions $\varphi_{j\sigma}(y)$ at the 
		end $y_\alpha$ of the wire, and these wave function form an 
		orthonormal set, the effective couplings are determined by the 
		BdG wave functions $u_{j\sigma}, v_{j\sigma}$ at the ends of the 
		wire. This is why we can relate the localization of the Majorana 
		wave functions to the couplings that determine the transmission 
		amplitude. 

	\subsection*{Coupling matrix elements and energy levels}
		To obtain the effective couplings $\lambda_{j\sigma\alpha}^e(\mu)$ 
		($\lambda_{j\sigma\alpha}^h(\mu)$) for electron and hole like 
		co-tunneling processes, we consider a tunneling event in which 
		the dot is initially in the state $\ket{N_0,\{n_i\}}$ and
		where co-tunneling takes place via an excited state 
		$\ket{N_0\pm1,\{n_i'\}}$. The couplings are  given by the overlap 
		$\bra{N_0,\{n_i\};\{\alpha,\sigma\}}H_{\rm tun}\ket{N_0\pm1,\{n_i'\}}$. In 
		principle, we would need to consider overlap of condensate wave 
		functions with different numbers of Cooper pairs, which however 
		are not easily accessible in the BdG formalism. We hence neglect 
		this contribution and use Eq.~\eqref{SEq:CouplSCDef} to determine 
		the couplings,  and we choose the chemical potential $\mu$ in the 
		computation of the wave functions such that it corresponds to 
		intermediate BdG state with $N_0\pm 1$ electrons through which 
		the tunneling occurs.\\
		We separately consider electron-like and hole-like processes and 
		distinguish between even and odd particle number in the ground 
		state. 
		As only pairs of electrons can enter the condensate 
		of the superconductor, the transmission depends on
		the number parity of $N_0$.
		For even $N_0$, the $T=0$ ground state is given by $\ket{N_0,\{n_i=0\};\{\alpha,\sigma\}}$, 
		i.e. all electrons are in the condensate.
		For odd $N_0$, we assume that one electron resides in the first BdG eigenstate such that the
		ground state is given by
		$\ket{N_0,\{n_1=1,n_{i\neq 1} = 0\};\{\alpha,\sigma\}}$.
		Electron-like  intermediate states are 
		$\ket{N_0+1,\{n_m'=1,n'_{i\neq m}=0\}}$ for even $N_0$ and 
			$\{\ket{N_0+1,\{n_i'=0\}},\,\ket{N_0+1,
			\{n_1'=1,n_m'=1,n'_{i\notin\{1,m\}}=0\} }\}$ 
		for odd $N_0$. For hole-like co-tunneling, 
		intermediate states have $N_0-1$ electrons and the occupancy of 
		bogolubons changes in an analogous way to the electron excited 
		states described above. In this way, we find the following 
		effective couplings for $T=0$
		\begin{align}
			\lambda^h_{\alpha, j,\sigma}(N_0,\{n_i\}) &=
			 \begin{cases} 
				\lambda^u_{\alpha,j,\sigma}(\mu_j^h(\{n_i\})) 
					& \text{for $n_j = 1$}\\
				\lambda^v_{\alpha,j,\sigma}(\mu_j^h(\{n_i\})) 
					& \text{for $n_j = 0$}
			 \end{cases}\\
			\lambda^e_{\alpha, j,\sigma}(N_0,\{n_i\}) &=
			 \begin{cases} 
				\lambda^v_{\alpha,j,\sigma}(\mu_j^e(\{n_i\})) 
					& \text{for $n_j = 1$}\\
				\lambda^u_{\alpha,j,\sigma}(\mu_j^e(\{n_i\})) 
					& \text{for $n_j = 0$}
			 \end{cases}
		\end{align}
		where the superscript $e$ denotes electron-like and $h$ hole-like 
		couplings, $\alpha$ is a lead index, $\sigma$ the spin of the 
		tunneling electron in the lead, and $n_j$ is the state through 
		which the tunneling occurs.
			
		In addition, we consider  excited initial states $\ket{N_0,\{n_i\}}$ 
		with energy $E(\{n_i\}) = \sum_{j, n_j = 1} \mathcal{E}_j$, whose 
		statistical weight is described by the Boltzmann factor 
		$\e^{-\beta E(\{n_i\})}$ with $\beta=1/k_BT$. For even $N_0$,  an even number of BdG 
		states has to be occupied, and  an odd number of BdG states for odd 
		$N_0$.	Numerically computing the excitation energies reveals that 
		excited states with occupied levels $n_{j>10}=1$ have too small 
		Boltzmann factors to significantly contribute to the transmission 
		amplitude. We therefore restrict ourselves to excited states where 
		only levels with energies among the ten smallest ones can be occupied.	
		Also states with more than three occupied levels yields negligible 
		weights and are therefore neglected. In addition, we do not recompute 
		the chemical potential for each excited state and instead use the 
		chemical potential of the respective $T=0$ state.
			
		In addition to the tunneling matrix elements, the effective energies 
		of the intermediate states are needed. We find 
		\begin{align}
			{\varepsilon}_{{\rm eff},j}^h &=
			 \begin{cases}
				+\mathcal{E}_j(\mu_j^h(\{n_i\}))-eV_g+E_c(N_0-1) 
					& \text{for $n_j = 1$}\\
				-\mathcal{E}_j(\mu_j^h(\{n_i\}))-eV_g+E_c(N_0-1) 	 
					& \text{for $n_j = 0$}
			 \end{cases}\\
			{\varepsilon}_{{\rm eff},j}^e &=
			 \begin{cases}
				-\mathcal{E}_j(\mu_j^e(\{n_i\}))-eV_g+E_cN_0 
					& \text{for $n_j = 1$}\\
				+\mathcal{E}_j(\mu_j^e(\{n_i\}))-eV_g+E_cN_0 	 
					& \text{for $n_j = 0$} \  . 
			 \end{cases}
		\end{align}
		Using these, 
		we obtain 
		the transmission amplitude from the scattering matrix Eq.~\eqref{Eq:Weid} 
		via $T_{\up\up}=S(1,3)$ and $T_{\down\down} = S(2,4)$.  
		In the thermal average, we define $Z 
		= \sum_{\{n_i\}|N_0} \e^{-\beta E(\{n_i\})}$ where number parity of 
		$\sum_i n_i$ is determined by the number parity of $N_0$.   
		The anti-unitary reflection symmetry $\tilde{\Pi} \varphi_j(y)=K\varphi_j(L-y)$ 
		(where $\varphi_j$ are eigenfunctions of $H_{\rm wire}$) ensures 
		that $T_{\sigma\sigma}$ is imaginary in the middle between resonances
		where the real part of the denominator $\approx E_c/2$ is large
		compared to the level broadening.
		We consider transmission through the first $j_{\rm max}$ levels.  
			
		\renewcommand{\theequation}{B\arabic{equation}}
		\setcounter{equation}{0}

		\section*{APPENDIX B: Analytic solution for Majorana wave function } \newcommand{\red}{\tilde}
		 \begin{figure}[t!]	
			\centering
			\includegraphics[width=8.6cm]{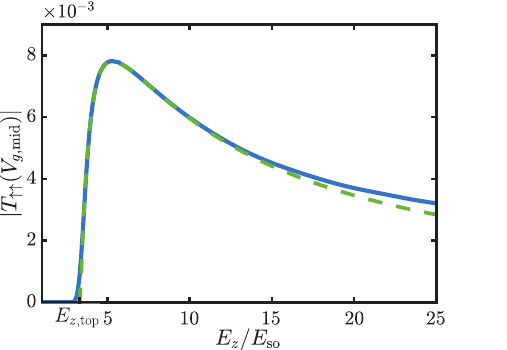}
			\caption{\label{Fig:SUPPAnalyticVsNumeric}  
			Comparison between numerical result (solid, blue) for the transmission 
			amplitude $|T_{\up\up}(V_{g,\rm mid})|$ in the topological region as a 
			function of the magnetic field for $j_{\rm max}=200$ level, $N_w=35$ and 
			$\Delta=2\,E_{\rm so}$ and full analytical solution (dashed, green), 
			in very good agreement with the numerical result. For very large Zeeman
		  energies, higher levels contribute to the transmission and the overlap 
			between the MZMs from the ends is finite such that there 
			is a small deviation between analytic and numeric results.}
		\end{figure}
		 
		 \begin{figure}[t!]	
			\centering
			\includegraphics[width=8.6cm]{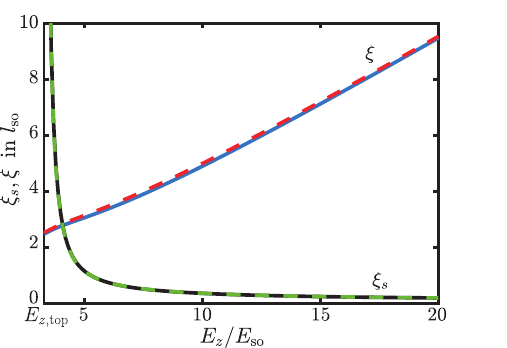}
			\caption{\label{Fig:FullVsApprox} Full analytic correlation lengths
				$ \xi _1=l_{\rm so}{\rm Re}(A_1)^{-1}$ and $ \xi _2=l_{\rm so}{\rm Re}(A_3)^{-1}$ (solid lines)
				and approximations $ \xi  $ and $ \xi_s$ (dashed lines). 
				Here, we used the self-consistent chemical potential $\mu(E_z)$
				for fixed particle number $N_w=35$ in the wire and $\Delta=2\,E_{\rm so}$.}
		\end{figure}	
		
		The BdG equations of a semi-infinite Rashba wire can in fact be solved analytically 
		for an exact zero energy state \citep{Sarma.2012}. Therefore, assuming a sufficiently 
		long wire such that the Majorana wave functions of both ends have negligible overlap, 
		one can derive an analytical expression for the Majorana wave functions. In this 
		section, we present the analytical solution and a series of approximations that help 
		to understand the emergence of the maximum of the transmission amplitude at the onset 
		of the topological regime.   
		
		The solution of the BdG equation $\mathcal{H}_{\rm wire}\psi = 0, \; \psi(0)=0$ for a 
		zero energy state with $\mathcal{H}_{\rm wire}$ defined in Eq.~\eqref{Eq:Hwire} has 
		the form $\psi=(\chi_{\uparrow},\chi_{\downarrow},i\chi_{\downarrow},i\chi_{\uparrow})$. 
		By making use of the real functions $\exp({i\pi/4})\chi_\up=\hat{\chi}_\up\in\mathbb{R}$, 
		$i\exp({i\pi/4})\chi_\down =\hat{\chi}_\down\in\mathbb{R}$ the BdG equation reduces to 
		the two equations
		\begin{align}
		 \begin{split}
			-\partial_{\red{y}}^2\hat{\chi}_\up - \red{\mu} \hat{\chi}_\up - \red{E_z}\hat{\chi}_\up - 2\partial_{\red{y}}\hat{\chi}_\down 
				+ \red{\Delta} \hat{\chi}_\down &= 0   \\
			-\partial_{\red{y}}^2\hat{\chi}_\down - \red{\mu} \hat{\chi}_\down + \red{E_z}\hat{\chi}_\down + 2\partial_{\red{y}}\hat{\chi}_\up 
				- \red{\Delta} \hat{\chi}_\up &= 0 \ .
		 \end{split} \label{Eq:RashbaAnaRedBdG}
		\end{align}
		Here, we use reduced quantities $\red{\mu}=\mu/E_{\rm so}$, $\red{E_z}=E_z/E_{\rm so}$,
		$\red{\Delta}=\Delta/E_{\rm so}$, $\red{y}=y/l_{\rm so}$, and $\red{k}=k l_{\rm so}$. 
		Since Majorana modes are expected to be exponentially localized at the end of the semi-infinite
		wire, we use an ansatz 
		\begin{align}
				\left(\begin{matrix} \hat{\chi}_\up \\ \hat{\chi}_\down \end{matrix}\right) 
				&= \e^{-A {\red{y}}} \left(\begin{matrix} \varrho_\up \\ \varrho_\down \end{matrix}\right) \ .
		\end{align}
		With this ansatz, we obtain the system of equations
		\begin{align}
			\left(\begin{matrix} -A^2-\red{\mu}-\red{E_z} & 2A+\red{\Delta} \\ -2A-\red{\Delta}& 
				-A^2-\red{\mu}+\red{E_z}  \end{matrix}\right)  \left(\begin{matrix} \varrho_\up \\ \varrho_\down \end{matrix}\right)  &= 0 \ .  
			\label{Eq:AnaRashAnsatzInDE}
		\end{align}
		The requirement for a non-trivial solution, i.e. a vanishing determinat of the coefficient matrix,
		yields the quartic equation
		\begin{align}
			 0  
				&= A^4 + A^2 (2\red{\mu}+4) + A (4\red{\Delta}) + \red{\mu}^2 - \red{E_z}^2 + \red{\Delta}^2  
		\end{align}
		in $A$ which is already in the reduced form $A^4+A^2 \alpha + A \beta +\gamma = 0$ and can be solved
		analytically. By factorizing the polynomial $0=(A-A_1)(A-A_2)(A-A_3)(A-A_4)$ using its four roots, and
		comparing to the above equation, one finds that $0=A_1+A_2+A_3+A_4$ and $A_1A_2A_3A_4=
		\red{\mu}^2-\red{E}_z^2+\red{\Delta}^2$. The four solutions are given by
		\begin{align}
			 A_i  &=  \frac{1}{2}\left[\pm_1 W\pm_2\sqrt{W^2-4(\alpha+Y\pm_1 Z)}\right] \ . \label{Eq:SolutionAi}
		\end{align}
		with the abbreviations $P=-\alpha^2/12 -\gamma$, $Q=-\alpha^3/108 +\alpha\gamma/3-\beta^2/8$, 
		$U=(-Q/2+\sqrt{Q^2/4+P^3/27})^{1/3}$, $Y=-5\alpha/6+U-P/(3U)$, $W=\sqrt{\alpha+2Y}$, and
		$Z=\beta/(2W)$. Here, $\pm_1$ and $\pm_2$ can individually be $+1$ or $-1$ to give rise to four solutions $A_i$.
		In the topological regime $\gamma=\red{\mu}^2 - \red{E_z}^2 + \red{\Delta}^2<0$,
		it can be shown that a solution with ${\rm Re} A_1,{\rm Re} A_2,{\rm Re} A_3>0$, $A_1=A_2^*$, ${\rm Im} A_3=0$,
		and ${\rm Re}A_4<0$ exists. To be able to normalize the solution, the coefficient of the $A_4$ term 
		needs to vanish. Then,  Eq.~\eqref{Eq:AnaRashAnsatzInDE} has the solution
		\begin{align}
					\left(\begin{matrix} {\varrho_{\up,i}} \\ {\varrho_{\down,i}} \end{matrix}\right) 
				&= \mathcal{N}_i\left(\begin{matrix} {2A_i+\red{\Delta}}\\{A_i^2+\red{\mu}+\red{E_z}}\end{matrix}\right)   \ ,
		\end{align}
		where $\mathcal{N}_i$ are normalization constants. In the topological regime, we define
		\begin{align}
			A_1 &= \red{\xi}_1^{-1} + i \red{k}_{\rm eff} \label{Eq:XiFromA1}\\
			A_2 &= \red{\xi}_1^{-1} - i \red{k}_{\rm eff}\\
			A_3 &= \red{\xi}_2^{-1}  \ .
		\end{align} 
		Therefore, the Majorana wave function 
		\begin{align}
			\hat{\bm{\chi}}_{\rm L}(y) &= \mathcal{N}\Bigg[
				\e^{-y/\xi_2}
				\left(\begin{matrix} {2\red{\xi}_2^{-1}+\red{\Delta}}\\{\red{\xi}_2^{-2}+\red{\mu}+\red{E_z}} \end{matrix}\right) \notag\\
				&
				+\e^{-y/\xi_1} \Bigg\{ 
					a\e^{i k_{\rm eff}y}
					\left(\begin{matrix} {2(\red{\xi}_1^{-1}+i \red{k}_{\rm eff})+\red{\Delta}}\\
						{(\red{\xi}_1^{-1}+i \red{k}_{\rm eff})^2 + \red{\mu}+ \red{E_z}}\end{matrix}\right) \notag\\
				&
					+b\e^{-i k_{\rm eff}y}
					\left(\begin{matrix} {2(\red{\xi}_1^{-1}-i \red{k}_{\rm eff})+\red{\Delta}}\\
						{(\red{\xi}_1^{-1}-i \red{k}_{\rm eff})^2 + \red{\mu}+ \red{E_z}}\end{matrix}\right)
				\Bigg\}\Bigg]\ . \label{Eq:RashbaAnaSol}
		\end{align} 
		consist of a evanescent term with localization length $\xi_2$
		and an oscillating term with localization length  of the envelop $\xi_1$. Here, the boundary
		condition $\hat{\bm{\chi}}_{\rm L}(0)=0 $ fixes the coefficients
		\begin{widetext}
		\begin{align}
			a = b^* = \frac
				{(i\red{\xi}_2+\red{\xi}_1(-i+\red{k}_{\rm eff}\red{\xi}_2))(-2+2(\red{E_z}+\red{\mu})
					\red{\xi}_1\red{\xi}_2-\red{\Delta}(\red{\xi}_1+\red{\xi}_2)+i \red{k}_{\rm eff}
					\red{\xi}_1(2+\red{\Delta}\red{\xi}_2))}
				{4\red{k}_{\rm eff}(1+\red{\xi}_1(\red{\Delta}-(\red{E_z}-\red{k}_{\rm eff}^2+\red{\mu})\red{\xi}_1))\red{\xi}_2^2}\ .
		\end{align}
		\end{widetext}
		In the following, we refer to this solution as  "analytic solution". 
		To compute the transmission amplitude, we use that the Majorana wave function at the right
		wire end is given by $\chi_{R}(y)\propto \chi_L^*(L-y)$ and evaluate the overlap with decaying
		wave functions from the leads.	
		
		We find that the analytic expression is in very good agreement with the numerical results (Fig.~\ref{Fig:SUPPAnalyticVsNumeric}) 
		in the 
		topological regime, even when taking transport through many levels into account. However, 
		without some approximations it is difficult to gain much insight into the lengthy analytical
		expression. In order to make progress, we first use that the oscillations of the Majorana wave functions is approximately 
		determined by the Fermi momentum
		\begin{align}
				k_{\rm eff} &\approx \red{k}_{F} = \sqrt{ 2+\red{\mu} + \sqrt{\red{E}_z^2+4+4\red{\mu}}} \ ,
		\end{align}
		which is nearly independent of $E_z$ in the case where the chemical potential is 
		self-consistently determined to fix the particle number in the wire.
		In addition, the localization length of the oscillating term can be approximated by the
		coherence length due to the p-wave gap
		\begin{align}
				\xi_1 &= {\rm Re}(A_1)^{-1}\,l_{\rm so} \approx  \xi = \frac{\hbar v_F}{\Delta_{p,\rm ind}} 
						  \ . 
		\end{align}
		The evanescent term however, has a different correlation length  whose divergence at the topological phase transition is governed  by the closing of the topological gap at $E_z = E_{z, \rm top}$
		\begin{align}
				\xi_2  = A_3^{-1} \,l_{\rm so} \approx \xi_s  
				&= \left( -\xi^{-1} + \sqrt{\xi^{-2} - \frac{\red{\mu}^2+\red{\Delta}^2-\red{E}_z^2}{\xi^{-2}+k_F^2}} \right)^{-1}
				\notag\\
				&\propto  \frac{1}{E_z-\sqrt{\Delta^2+\mu^2}}  \ . 
		\end{align}
		Here, we used the relations between the $A_i$ above Eq.~\eqref{Eq:SolutionAi} to express 
		$A_3$ in terms of $A_1, A_2$ and ultimately in terms of $\xi$. 		
		We find that the approximations for the localization lengths 
		(dashed lines in Fig.~\ref{Fig:FullVsApprox}) are in excellent agreement with
		the exact analytical expressions (solid lines in Fig.~\ref{Fig:FullVsApprox}) 
		in the whole topological regime. For the approximation in the main text, the Majorana wave functions are 
		reduced to the sum of the envelops of oscillating and evanescent term,
		neglecting the spin dependence and the oscillations. As the couplings are 
		determined by the wave function weights at the ends of the wire, the 
		oscillations are less important. However, in the evanescent term, the spin-$\downarrow$
		component can be larger than the spin-$\uparrow$ component at the beginning 
		of the topological regime.		
		Nevertheless, this rough approximation is still in good agreement with the numerical results
		for the case where $k_F$ is approximately constant (as for a fixed particle number
		in the wire) and allows to understand the occurrence of the amplitude maximum.

%

\end{document}